\documentclass[11pt,a4paper]{article}
\usepackage[utf8]{inputenc}
\usepackage{amsmath}
\usepackage{amsfonts}
\usepackage{amssymb}
\usepackage{array,multirow}
\begin{document}


\centerline{\bf{Fluctuation-Dissipation relation from anomalous stress tensor and Hawking effect}}
\begin{center}
Rabin Banerjee$^a$ and Bibhas Ranjan Majhi$^b$\\
$^a$S. N. Bose National centre for Basic Sciences, JD Block, Sector III, Salt Lake, Kolkata-700098, India\\
	$^b$Department of Physics, Indian Institute of Technology Guwahati, Guwahati 781039, Assam, India\\
e-mails: rabin@bose.res.in, bibhas.majhi@iitg.ac.in
\end{center}

\begin{abstract}
We show a direct connection between Kubo's fluctuation-dissipation relation and Hawking effect that is valid in {\it any dimensions for any stationary or static black hole}. The relevant correlators corresponding to the fluctuating part of the force, computed from the known expressions for the {\it anomalous} stress tensor related to gravitational anomalies, are shown to satisfy the Kubo relation, from which the temperature of a black hole as seen by an observer at an arbitrary distance is abstracted. This reproduces the Tolman temperature and hence the Hawking temperature as that measured by an observer at infinity.
\end{abstract}

\section{Introduction}

Quantisation of fields on a classical background, containing a horizon, leads to the fact that particles can radiate from the horizon \cite{Hawking:1974rv,  Unruh:1973,  Takagi:1986kn}. Moreover, it is now well known that such phenomenon depends on a particular observer and is connected to the non-unique definition of vacuum in a non-inertial frame or in curved spacetime. One of the classic examples in this context is  the Hawking effect \cite{Hawking:1974rv} --  the thermal radiation observed by a static observer at infinity in a black hole spacetime in the Kruskal or Unruh vacuum. While there are several approaches to analyse this phenomenon, each has its own merits and/or demerits, but none is truly clinching. This is the primary reason that new avenues are still being explored. 

The fluctuation–dissipation theorem is a general result of statistical thermodynamics that yields a concrete relation between the fluctuations in a system that obeys detailed balance and the response of the system to applied perturbations (see \cite{Kubo} for a review). It has been effectively used to study various processes like Brownian motion in fluids, Johnson noise in electrical conductors and, as shall become clear very soon, relevantly, Kirchoff's law of thermal radiation. Since black holes satisfy the condition of detailed balance and the emitted radiation is thermal, it is likely that new insights into the phenomenon of Hawking radiation could be obtained by using the fluctuation-dissipation relation.

 Our motivation in this paper is to exploit the fluctuation-dissipation relation to obtain the Hawking temperature. In fact we are able to obtain the more general Tolman temperature \cite{ Tolman:1930zza, Tolman:1930ona}, that is valid for an observer at an arbitrary distance. There are different versions of the theorem but the one suited for our analysis was given by Kubo \cite{Kubo,  kubo_book,  Reif}. In simple terms this relation is able to provide the temperature of the heat bath from a study of the correlators of the fluctuations of the force of the emitted particles, as measured by the detector. We shall apply this relation to black holes. Treating the black holes as a heat bath, it is possible to compute the fluctuations of the force of the emitted particles as seen by an observer at an arbitrary distance. Using Kubo's relation we derive the temperature, which turns out to be the Tolman temperature. Putting the detector at infinity immediately yields the Hawking temperature \cite{Hawking:1974rv}. 
 
An essential ingredient in our calculation is the structure of the energy momentum tensor. The force is computed by taking the time variation of the space component of the four momentum which in turn is defined from the energy momentum tensor. Since our analysis is very near the horizon, where the spacetime (arbitrary dimensional static as well as stationary black holes) is effectively $(1+1)$ dimensional \cite{Robinson:2005pd,  Iso:2006wa,  Majhi:2011yi}, we  shall concentrate on the two dimensional stress tensor in a curved background. Classically this is not well defined and recourse has to be taken to some regularisation to include quantum effects. Incidentally the method discussed  here was applied earlier by one of the authors, in a collaborative work \cite{Adhikari:2017gyb,  Das:2019aii}, in a classical treatment. It is useful to recall that the fluctuation dissipation relation remains valid both for classical and quantum systems. Other relevant applications were based on the path integral fluctuation-dissipation formalism developed in \cite{Caldeira}. The Minkowski vacuum
	was modelled as a thermal bath with respect to an accelerated observer,
	so that any particle in it is executing a Brownian like motion \cite{Unruh:1989, Raine:1991, Hinterleitner:1993, Kim:1997, Kim:1998}. For the quantum case which is relevant here, there are two possibilities. Either the theory is non-chiral or it is chiral. In the first case the stress tensor satisfies diffeomorphism invariance but lacks conformal invariance leading to a nonvanishing trace of the stress tensor \cite{Deser:1976yx,  Duff:1977ay,  Polyakov:1981rd, Christensen:1977jc}. For the chiral theory both conformal and diffeomorphism symmetries are broken \cite{Bardeen:1984pm, AlvarezGaume:1984dr, AlvarezGaume:1983ig, Leutwyler:1984nd}. We have done the analysis for either situation  and find the correct Tolman/Hawking temperatures. 
	
	There are certain differences and important aspects of the present analysis compared to the earlier related investigations. As the analysis is done very near the horizon where the effective metric is $(1+1)$ dimensional, the appearance of the gravitational anomalies (trace and/or diffeomorphism), at the quantum level, is unavoidable. Since our calculation is at the quantum regime, dealing with corresponding anomalous stress tensor is quite natural which was lacking in literature. This is much more consistent with the quantum nature of black holes. Here we shall do exactly the same to find the fluctuating part of the force as experienced by a relevant observer. We shall show that this satisfies the equilibrium fluctuation-dissipation relation. Furthermore, near horizon effective $(1+1)$ geometry is very much general for not only static black holes, but also for stationary ones in any arbitrary dimensions. Therefore our results are valid beyond the two dimensional static black hole spacetimes as far as near horizon physics is concerned.
All these points are new in showing, in very general terms outlined above, the connection between the fluctuation-dissipation theorem and particle production in a background black hole metric.
It shows that as far as near horizon is concerned, the standard form of fluctuation-dissipation relation is very universal. Moreover, in our analysis the concept of temperature is emerging by comparing this with  Kubo’s form of fluctuation-dissipation relation, which is given by Tolman's expression; instead of looking at the emission spectrum of particles. This can be viewed as an alternative way of understanding the thermality of the horizon. So it presents a much more general aspect of particle production in black hole spacetimes.

The organisation of the paper is as follows. In the next section, we shall briefly discuss our general formalism of computing fluctuations of force and their use in Kubo's relation. In sections \ref{Sec3} and \ref{Sec4}, respectively, we derive the Hawking effect from the non-chiral and chiral theories using the corresponding anomalous stress tensor. The last section contains our conclusions and a look into future possibilities.

\section{Fluctuations of force and Kubo's relation}
\label{Force}

Here we outline our general strategy for obtaining the Hawking temperature from Kubo's relation. The first thing is to define the force that will lead to the force correlators. The space components of the four momentum as measured by the observer in its own frame which perceives the particles is defined as
\begin{equation}
p^\alpha=\int d^3{\bf{x}}\sqrt{h}~T^{0\alpha}(\tau,{\bf{x}})\frac{\delta^{(3)}({\bf{x}}-{\bf{x}}_D)}{\sqrt{h}} ~,
\label{3.01n}
\end{equation}
where $h$ is the determinant of the induced metric of the $t=$ constant hypersurface for the detector's metric and ${\bf{x}}_D$ is the position of the detector in its own frame. Here $\sqrt{h}$ has been introduced both with volume element and the Dirac-delta function to define the most general invariant quantities which remain same under a general coordinate transformation for a general space. 
Then Eq.  (\ref{3.01n}) yields
\begin{equation}
p^\alpha = T^{0\alpha}(\tau,{\bf{x}}_D)~.
\label{3.01}
\end{equation}
Note that ${\bf x}_D$ is not changing. Therefore, the above quantity must be a function of its proper time $\tau$ only. Then the force, as measured by this detector, turns out to be
\begin{equation}
F^\alpha(\tau)=\frac{dp^{\alpha}}{d\tau} = \frac{dT^{0\alpha}}{d\tau}~.
\label{3.02}
\end{equation} 
The fluctuating part of this force is
\begin{equation}
F^\alpha_{fluc} (\tau)=F^\alpha-<F^\alpha> = \frac{dT^{0\alpha}}{d\tau} - \frac{d}{d\tau}<T^{0\alpha}>~.
\label{3.03}
\end{equation}
Next we define the fluctuating force-force two point correlation function as
\begin{equation}
R^{\alpha\beta} (\tau_2;\tau_1) = <F^{\alpha}_{fluc}(\tau_2)F^{\beta}_{fluc}(\tau_1)>~.
\label{3.04}
\end{equation}
In the above $<\dots>$ refers to the expectation value of the  quantity of interest with respect to the relevant state. It is now possible to proceed with the calculations in a general way. More precisely, for the accelerated frame case, the stress tensor components are defined in the Rindler frame while the states are vacuum states corresponding to the Minkowski observer. 

Specialising   for black holes, the operators are defined in static Schwarzschild coordinates and the states will be considered as Kruskal and Unruh vacua. Moreover, near the horizon, the arbitrary dimensional static and stationary black holes are effectively two dimensional and hence conformally flat \cite{Robinson:2005pd,  Iso:2006wa,  Majhi:2011yi}. 
The idea is the following. Consider for simplicity, say a massive scalar field, in the original stationary background. The explicit expansion of the field action for this black hole metric in the near horizon limit, upon integrating out the angular coordinates, reduces to an infinite set of two dimensional theories enumerated by angular and magnetic quantum numbers. One can check that the dimentionally reduced form of the action is similar to that of a collection of infinite free massless scalar field modes, each of which is residing on an effective ($1+1$) dimensional conformally flat metric. This has been checked case by case. For example, see \cite{Robinson:2005pd} in the case of static spherically symmetric black hole (also see section $9.3$ of \cite{Frolov1} and for higher dimensional black holes look at section $9.10$ of the same). The (1+3) dimensional Kerr-Newman and other higher dimensional rotating cases (e.g. Myers-Perry black hole) have been investigated as well \cite{Iso:2006wa,Umetsu:2009ra,Banerjee:2010ye}. In all instances, the general form of the effective metric, in Schwarzschild coordinates, is given by
\begin{equation}
ds^2= -f(r)dt^2+\frac{dr^2}{f(r)}~.
\label{ref1}
\end{equation}
The explicit value of $f(r)$ is different for different black holes. For instance, in the case of the Kerr-Newman metric this is given by \cite{Umetsu:2009ra}
\begin{equation}
f(r)= \frac{r^2+a^2-2Mr+Q^2}{r^2+a^2}~,
\end{equation}
where $M, Q,$ and $a$ are mass, charge and angular momentum per unit mass of the black hole, respectively. Since we investigate the thermal behaviour of horizons, which is dominated by the near horizon physics, the above form of the metric is sufficient as far as near horizon phenomenon is concerned. Moreover we shall see that the explicit form of the metric coefficient is not necessary for our purpose, and in this sense the results, obtained using (\ref{ref1}),  as far near horizon is concerned, will be valid for any stationary black hole.
The above effective metric in  Kruskal null-null coordinates and in Eddington null-null coordinates is given by
\begin{eqnarray}
ds^2&=& \frac{f(U,V)}{\kappa^2 UV} dUdV; ~~ {\textrm{Kruskal null-null}}
\nonumber
\\
&=&-f(u,v)dudv; ~~ {\textrm{Eddington null-null}}~,
\label{3.15}
\end{eqnarray}
where $f$ is the metric coefficient and $\kappa =f'(r_H)/2$ is the surface gravity and $r=r_H$ is the horizon. The explicit form of $f(r)$ determines which higher dimensional metric is effectively given by the above near horizon form. For our present analysis, this explicit expression is not needed and thereby helps us to discuss all black holes (static as well as stationary) in a unified manner. The relations among these sets of coordinates are as follows:
\begin{eqnarray}
&&dr^*=\frac{dr}{f(r)}; \,\,\ u=t-r^*; \,\,\,\ v=t+r^*;
\\
&& U=-\frac{1}{\kappa}e^{-\kappa u}; \,\,\ \,\ V=\frac{1}{\kappa}e^{\kappa v}~.
\label{3.16}
\end{eqnarray}
Thus in Eq. (\ref{3.04}) we have to take only the $R^{11}$ component. Also, since the correlators are translationally invariant, it is a function of only the difference of the proper times, $(\tau_2-\tau_1)$. Let us now represent the Fourier transform of the correlator $R^{11}(\tau_2-\tau_1)$ by $K(\omega)$.

Now as usual \cite{Kubo, kubo_book, Reif}, the symmetric and antisymmetric combinations are taken as
\begin{equation}
K^\pm(\omega) = K(\omega)\pm K(-\omega)
\label{4.05}
\end{equation}
Then Kubo's fluctuation dissipation relation states that these two are related by,
\begin{equation}
K^+(\omega) = \coth\Big(\frac{\omega}{2T}\Big) K^-(\omega)~,
\label{4.07}
\end{equation}
where $T$ is the temperature of the heat bath which, in our case, is the black hole.
In the next sections we will explicitly compute the correlators, both for nonchiral and chiral cases,  and obtain the temperature from the relation (\ref{4.07}).

\section{\label{Sec3}Non-chiral theory}

At the quantum level both the trace and the covariant divergence of the stress tensor cannot be made vanishing. Since diffeomorphism symmetry is more fundamental in gravitational theories, a regularisation is done such that the trace of energy-momentum tensor is non-vanishing and given by \cite{Deser:1976yx, Duff:1977ay}: 
$T^a_a=c_w R$.
However, it is covariantly conserved, i.e. $\nabla_aT^{ab}=0$. The value of the proportionality constant is $c_w=1/(24\pi)$. In the (trace) anomaly based approach of discussing the Hawking effect \cite{Christensen:1977jc}, which is valid only for two dimensions, use is made of this result. However we are interested in the explicit form of the stress tensor. This is derived from  the anomalous effective action \cite{Polyakov:1981rd},
\begin{equation}
S_P =  -\frac{c_w}{4}\int d^2x \sqrt{-g}(-\phi\Box\phi+2R\phi)~,
\label{2.03}
\end{equation}
where,
\begin{equation}
\Box\phi=R~.
\label{2.04}
\end{equation}
By taking appropriate functional derivatives \cite{Polyakov:1981rd},
\begin{eqnarray}
T_{ab}&=&\frac{2}{\sqrt{-g}}\frac{\delta S_P}{\delta g^{ab}}=\frac{c_w}{2}\Big[\nabla_a\phi\nabla_b\phi-2\nabla_a\nabla_b\phi
\nonumber
\\
&+&g_{ab}(2R-\frac{1}{2}\nabla_c\phi\nabla^c\phi)\Big]~.
\label{2.05}
\end{eqnarray}

Now the equation for the scalar field (\ref{2.04}) under the background (\ref{3.15}) yields
\begin{equation}
\frac{\partial^2\phi}{\partial t^2}-\frac{\partial^2\phi}{\partial r^{*^2}} = -fR~.
\label{3.17}
\end{equation}
Since the metric is static, we choose the ansatz for the solution as
$\phi(t,r^*)=e^{-i\omega t}F(r^*)$,
where $\omega$ is the energy of the scalar mode and $F(r^*)$ is an unknown function to be determined. Substituting this in (\ref{3.17}) we obtain
\begin{equation}
\frac{d^2F}{dr^{*^2}} +\omega^2F = fRe^{i\omega t}~.
\label{3.19}
\end{equation}
Since our analysis is very near the horizon where $R$ is finite, the right hand side of the above can be neglected compared to the terms on the left hand side and then the solutions for $F$ are
$F=e^{\pm i\omega r^*}$.	
Under this limit,  the modes are identical to the usual ones and hence the mode expansion of $\phi$ is same as that for a free massless scalar field. Therefore the positive frequency Wightman functions, corresponding to respective vacuum states, will be same as those for the free massless scalar field. So the expressions for the positive frequency Wightman functions corresponding to Kruskal and Unruh vacuums are as follows \cite{book2}:
\begin{eqnarray}
&&G^+_K(x_2;x_1) = -\frac{1}{4\pi}\ln[(\Delta U - i\epsilon)(\Delta V -i\epsilon)]~;
\label{3.212}
\\
&&G^+_U(x_2;x_1) = -\frac{1}{4\pi}\ln[(\Delta U - i\epsilon)(\Delta v -i\epsilon)]~.
\label{3.21}
\end{eqnarray}

{\subsection{Kruskal vacuum}}
For Kruskal vacuum the expectation value of $T^{tr}$, as measured by the Schwarzschild static observer, vanishes (see Appendix C of \cite{Das:2019aii} for details). So the fluctuating part of the force is given by the first term of (\ref{3.03}). Therefore, the fluctuating force-force correlator, as measured by the Schwarzschild static observer, is given by 
\begin{eqnarray}
&&R^{11}_K(\tau_2;\tau_1)=\frac{d}{d\tau_2}\frac{d}{d\tau_1}<T^{tr}(\tau_2)T^{tr}(\tau_1)>
\nonumber
\\
&&=\frac{f^2(r_s)}{16}\frac{d}{d\tau_2}\frac{d}{d\tau_1}\Big[\Big(g^{uv}(\tau_2)\Big)^2 \Big(g^{uv}(\tau_1)\Big)^2 e^{-2\kappa(u_2+u_1)}
\nonumber
\\
&&\times<T_{UU}(\tau_2)T_{UU}(\tau_1)>\Big]
\nonumber
\\
&=&[f(r_s)]^{-2}\frac{d}{d\tau_2}\frac{d}{d\tau_1}\Big[e^{-2\kappa(u_2+u_1)}<T_{UU}(\tau_2)T_{UU}(\tau_1)>\Big]~,
\label{3.22}
\end{eqnarray} 
where $g^{uv} = -(2/f(r))$ has been used while the observer is static at $r=r_s$. In going from the first to the second equality, only the $T_{uu}$ (i.e. $T_{UU}$) component was considered. This is because this component, which corresponds to outgoing modes, leads to the flux of emitted particles from the horizon; while $T_{vv}$, related to ingoing modes, does not contribute to this flux.
Now using (\ref{2.05}) one finds
\begin{equation}
T_{UU}(x) =
\frac{c_w}{2}\Big[(\partial_U\phi)(\partial_U\phi)-2\partial^2_U\phi+\frac{2}{A}(\partial_UA)(\partial_U\phi)\Big]~,
\label{3.23}
\end{equation} 
where $A=f(U,V)/(UV)$. 
With this, one finds by using Wick's theorem
\begin{eqnarray}
&&<T_{UU}(x_2)T_{UU}(x_1)> = \Big(\frac{c_w}{2}\Big)^2\Big[ 4\partial^2_2\partial^2_1G(x_2;x_1)
+ 2\Big(\partial_2\partial_1G(x_2;x_1)\Big)^2
\nonumber
\\
&-& \frac{4\partial_1A_1}{A_1}\partial_2^2\partial_1G(x_2;x_1)
-\frac{4\partial_2A_2}{A_2}\partial_2\partial_1^2G(x_2;x_1)
\nonumber
\\
&+&\frac{4\partial_2A_2\partial_1A_1}{A_2A_1}\partial_2\partial_1G(x_2;x_1)\Big]~.
\label{3.09}
\end{eqnarray}
In the above we used the following notations: $\partial_i\equiv\partial_{U_i}$  and $\phi_i \equiv \phi(x_i)$ with $i=1,2$.
The expectation value is taken here with respect to the Kruskal vacuum. 
Other terms vanish as $<\phi>=0$.
In the above we denoted $<\phi_2\phi_1> = G(x_2;x_1)$ which is Green's function corresponding to the differential equation (\ref{2.04}) for field $\phi$. Here for Kruskal vacuum, this is given by (\ref{3.212}).
Substituting this we find the terms of the expression (\ref{3.09}) as
\begin{eqnarray}
&&\Big(\partial_2\partial_1G(x_2;x_1)\Big)^2=\frac{1}{16\pi^2}\frac{1}{(U_2-U_1)^4}~;
\nonumber
\\
&&\partial^2_2\partial^2_1G(x_2;x_1)=\frac{3}{2\pi}\frac{1}{(U_2-U_1)^4}~;
\nonumber
\\
&&\partial_2^2\partial_1G(x_2;x_1)=\frac{1}{2\pi}\frac{1}{(U_2-U_1)^3}~;
\nonumber
\\
&&\partial_2\partial_1^2G(x_2;x_1)=-\frac{1}{2\pi}\frac{1}{(U_2-U_1)^3}~;
\nonumber
\\
&&\partial_2\partial_1G(x_2;x_1)=-\frac{1}{4\pi}\frac{1}{(U_2-U_1)^2}~;
\label{3.12}
\end{eqnarray}
Hence the correlator (\ref{3.22}) turns out to be
\begin{eqnarray}
R^{11}_K(\tau_2;\tau_1) &=& - 16\kappa^6[f(r_s)]^{-3}\Big(\frac{c_w}{2}\Big)^2\Big(\frac{1}{8\pi^2}+\frac{6}{\pi}\Big)
\frac{5+4\sinh^2(\frac{\kappa}{2\sqrt{f(r_s)}}\Delta\tau)}{\sinh^6(\frac{\kappa}{2\sqrt{f(r_s)}}\Delta\tau)}
\nonumber
\\
&+&\frac{2\kappa^6[f(r_s)]^{-3}\Big[(\frac{f'(r_s)}{2\kappa})^2-1\Big]}{\pi}\Big(\frac{c_w}{2}\Big)^2\frac{3+2\sinh^2(\frac{\kappa}{2\sqrt{f(r_s)}}\Delta\tau)}{\sinh^4(\frac{\kappa}{2\sqrt{f(r_s)}}\Delta\tau)}~;
\label{3.25}
\end{eqnarray}
where we have used the transformation $U=-(1/\kappa)e^{-\kappa u}$, along with $\partial_UA/A=(1/U)((f'(r)/2\kappa)-1)$ while a  prime denotes  differentiation with respect to the $r$ coordinate and $\Delta u = u_2-u_1 = (t_2-r^*_s) - (t_1-r^*_s) = \Delta t = \Delta\tau/\sqrt{f(r_s)}$.

\vskip 3mm

{\subsection{Unruh vacuum}}
The expression for $R^{11}_U(\tau_2;\tau_1)$ is again given by (\ref{3.22}) where the vacuum expectation has to be calculated with respect to Unruh vacuum. This is because $<T^{tr}>$  for a Schwarzschild static observer is constant (see Appendix C of \cite{Das:2019aii} for a detailed analysis) and hence the second part of (\ref{3.03}) vanishes. The two point correlation function for the stress tensor component in this expression is again  expressed in terms of positive frequency Wightman function, which is given by (\ref{3.21}). Since the derivatives will be with respect to $U$, only the $\Delta U$ part of $G^+$ contributes and hence the final expression for $R^{11}_U(\tau_2;\tau_1)$ comes out to be the same as that of Kruskal vacuum; i.e. Eq. (\ref{3.25}).

\vskip 5mm
\section{\label{Sec4}Chiral theory}
Contrary to the previous nonchiral case, here both trace and diffeomorphism anomalies  exist:
$T^a_a = \frac{c_w}{2}R$;  $\nabla_bT^{ab}=\frac{c_w}{4}\bar{\epsilon}^{ac}\nabla_cR$.
This is the covariant form of the anomaly which, as was shown by \cite{Banerjee:2007qs,  Banerjee:2007uc,  Banerjee:2008wq,  Banerjee:2008sn}, is more effective than the consistent form of the anomaly, in analysing Hawking effect. Here the effective action and corresponding energy-momentum tensor are evaluated in \cite{Leutwyler:1984nd}. The form of the stress tensor is given by,
\begin{equation}
T_{ab} = \frac{c_w}{2}\Big[D_a G D_bG -2D_aD_bG+\frac{1}{2}g_{ab}R\Big]~,
\label{2.07}
\end{equation}
where $G$ satisfies $\Box G=R$. The chiral derivative is defined as $D_a = \nabla_a\pm\bar{\epsilon}_{ab}\nabla^b$. Here $+(-)$ corresponds to ingoing (outgoing) mode and $\bar{\epsilon}_{ab} = \sqrt{-g}\epsilon_{ab}$ (or $\bar{\epsilon}^{ab} = -\frac{\epsilon^{ab}}{\sqrt{-g}}$) is an anti-symmetric tensor while $\epsilon_{ab}$ is the usual Levi-Civita symbol in $(1+1)$ dimensions.

Since we are interested in outgoing modes, the negative sign of $D_a$ operator will be considered. In this case
then we have $D_U=2\nabla_U$ and $D_V=0$. This can be checked using the expression for  $\bar{\epsilon}_{ab} = \sqrt{-g}\epsilon_{ab}$ with $\epsilon_{UV}=1$. Then $T_{UU}$, as obtained from (\ref{2.07}), becomes  identical to the non-chiral case. Moreover, $G$ satisfies an equation which is identical to that satisfied by $\phi$. So the correlator for the fluctuation of the force, in both vacua, will be  identical to the form (\ref{3.25}). Only the over all multiplicative constant factor is different.

Note that in all cases, the form of the correlators for the fluctuation of the force are identical:
\begin{equation}
R^{11}(\tau_2;\tau_1) = C\frac{5+4\sinh^2(\frac{B}{2}\Delta\tau)}{\sinh^6(\frac{B}{2}\Delta\tau)}+C_0\frac{3+2\sinh^2(\frac{\kappa}{2\sqrt{f(r_s)}}\Delta\tau)}{\sinh^4(\frac{\kappa}{2\sqrt{f(r_s)}}\Delta\tau)}~;
\label{3.26}
\end{equation} 
where $C$ and $C_0$ are unimportant constants (which can be different for different cases) and $B$ is given by,
\begin{equation}
 B=\kappa/\sqrt{f(r_s)}~.
 \label{newequation}
 \end{equation}
Moreover as the correlator depends only on the difference of the detector's proper time, it is time translational invariant. This is a  signature of the thermal equilibrium between the detector and the thermal bath seen by this detector. It also helps us to express the quantity in its Fourier space which we shall do later to find the equilibrium fluctuation-dissipation relation.

 The Fourier transformation of (\ref{3.26}) is given by
\begin{eqnarray}
&&K(\omega) 
= C \int_{-\infty}^{+\infty} d(\Delta\tau) \frac{5e^{i\omega\Delta\tau}}{\sinh^6(\frac{B\Delta\tau}{2}-i\epsilon)}
\nonumber
\\
&+& C\int_{-\infty}^{+\infty} d(\Delta\tau)\frac{4e^{i\omega\Delta\tau}}{\sinh^4(\frac{B\Delta\tau}{2}-i\epsilon)}
\nonumber
\\
&+&C_0\int_{-\infty}^{+\infty} d(\Delta\tau)\frac{3e^{i\omega\Delta\tau}}{\sinh^4(\frac{B\Delta\tau}{2}-i\epsilon)}
\nonumber
\\
&+&C_0\int_{-\infty}^{+\infty} d(\Delta\tau)\frac{2e^{i\omega\Delta\tau}}{\sinh^2(\frac{B\Delta\tau}{2}-i\epsilon)}~. 
\label{4.01}
\end{eqnarray}
The above equation involves four integrations. All of them can be evaluated by the standard formula \cite{Padmanabhan:1987rq}
\begin{eqnarray}
&&\int_{-\infty}^{+\infty} dx\frac{e^{-i\rho x}}{\sinh^{2n}(x-i\epsilon)} 
\nonumber\\
&=& \frac{(-1)^n}{(2n-1)!}\Big(\frac{2\pi}{\rho}\Big)\frac{1}{e^{\pi\rho}-1}\prod_{k=1}^{n}\Big[\rho^2+4(n-k)^2\Big].
\label{4.02}
\end{eqnarray}
Then one finds
\begin{equation}
K(\omega) = \Big(\frac{2\omega}{B}\Big)^3
\Big[\Big\{\Big(\frac{2\omega}{B}\Big)^2+4\Big\}\frac{\pi C}{6B}-\frac{2\pi C_0}{B}\Big]\frac{1}{e^{-\frac{2\pi\omega}{B}}-1}~.
\label{4.04}
\end{equation}
Now use of (\ref{4.05}) yields
\begin{equation}
K^+(\omega) = \coth\Big(\frac{\pi\omega}{B}\Big) K^-(\omega)~,
\label{4.08}
\end{equation}
which is identical to Kubo's fluctuation-dissipation relation (\ref{4.07})
with the temperature identified as
\begin{equation}
T=\frac{B}{2\pi}~.
\label{4.09}
\end{equation} 
Using (\ref{newequation}) one can check that this corresponds to,
\begin{equation} 
T =  \kappa/(2\pi\sqrt{f(r_s)})
\label{tolman}
\end{equation}
which is the correct value of the Tolman expression \cite{Tolman:1930ona, Tolman:1930zza}. For the detector located at infinity, $r_s\rightarrow\infty$, $f(r_s)\rightarrow 1$, the above result simplifies to,
\begin{equation} 
T =  \kappa/2\pi
\label{hawking}
\end{equation}
which is the familiar Hawking expression \cite{Hawking:1974rv}.

\section{Conclusions}
A new approach for analysing the Hawking effect has been given in this paper which is based on the fluctuation dissipation relation as formulated by Kubo. It is general enough to include any stationary metric, any dimensions and also to yield the Tolman temperature, which is the result of measurement by an observer at an arbitrary distance from the black hole horizon. Expectedly, the result for an observer at infinity is easily derived, thereby giving the Hawking temperature. It is universal in the sense that it does not depend on how the effective action yielding the stress tensor is regularised. Thus it was applicable both for  nonchiral and chiral couplings. In the literature \cite{Christensen:1977jc, Robinson:2005pd,  Iso:2006wa, Banerjee:2007qs,  Banerjee:2007uc,  Banerjee:2008wq,  Banerjee:2008sn}, stress tensor based approaches have used either one or the other but a holistic treatment was lacking. As mentioned in the introduction, the present analysis is much more improved and general as it dealt with the full quantum theory in the choice of the stress tensor and extends to any arbitrary dimensional stationary black hole. Therefore the present analysis provides a much more robust impact in the paradigm of thermality of the black hole horizon.

Contrary to several other approaches, this is a physically motivated derivation of the Hawking effect  by directly computing the force of the emitted spectrum on the detector and brings it in line with other phenomena of statistical thermodynamics. For instance, the present approach shows that the thermal heat bath characterising a black hole is Brownian in nature. Naturally such an approach is expected to yield further insights into the interpretation of black holes as thermodynamic objects.

It is possible to extend this analysis in other ways. As an example, the back reaction effect might be taken into account. This would change the force (and its fluctuations) as perceived by the detector. Application of the Kubo relation would then yield a correction to the Hawking temperature that determines the greyness of the otherwise black body radiation. 

Finally, we mention that our present analysis is general in the sense that all the static as well as stationary solutions of Einstein's gravity (with or without cosmological constant) have been addressed. Moreover the calculation has been done close to an event horizon. In this regard, it may be mentioned that there are spacetimes (e.g. see \cite{Frolov2}), which possess other types of horizon; e.g. Killing horizon, conformal Killing horizon, apparent horizon etc. It has been observed that they can also emit at the quantum level (some of the cases are discussed in \cite{Frolov2}). Therefore it would be interesting to investigate such examples under the present scenario. In that case one should first check if the near horizon is effectively ($1+1$) dimensional. If so, then only the use of two dimensional gravitational anomalies can be done. In this case, it would be interesting to check the fate of the Tolman expression. All these need a thorough and detailed investigation which is beyond the scope of the present paper. Therefore we leave these cases for our future studies.


\begin{thebibliography}{99}
	\bibitem{Hawking:1974rv} 
	S.~W.~Hawking,
	``Black hole explosions,''
	Nature {\bf 248}, 30 (1974).
	
	\bibitem{Unruh:1973} 
	W.~G.~Unruh,
	``Notes on black hole evaporation,''
	Phys.\ Rev.\ D {\bf 14}, 870 (1976).
	
	\bibitem{Takagi:1986kn} 
	S.~Takagi,
	``Vacuum noise and stress induced by uniform accelerator: Hawking-Unruh effect in Rindler manifold of arbitrary dimensions,''
	Prog.\ Theor.\ Phys.\ Suppl.\  {\bf 88}, 1 (1986).  
	
	\bibitem{Kubo}
	R.~Kubo, 
	``The fluctuation-dissipation theorem,'' 
	Rep.\ Prog.\ Phys.\ {\bf 29}, 255 (1966). 
	
        \bibitem{Tolman:1930zza} 
	R.~C.~Tolman,
	``On the Weight of Heat and Thermal Equilibrium in General Relativity,''
	Phys.\ Rev.\  {\bf 35}, 904 (1930).
	
	\bibitem{Tolman:1930ona} 
	R.~Tolman and P.~Ehrenfest,
	``Temperature Equilibrium in a Static Gravitational Field,''
	Phys.\ Rev.\  {\bf 36}, no. 12, 1791 (1930).
	
	\bibitem{kubo_book}
	R.~Kubo, M.~Toda and N.~Hashitsume, 
	``Statistical Physics II:
	Nonequilibrium Statistical Mechanics,''  Springer-Verlag,
	Berlin Heidelberg, (1985). 
	
	\bibitem{Reif}
	F.~Reif, 
	``Fundamentals of Statistical and Thermal Physics,''
	McGraw-Hill, New York, (1965).
	
\bibitem{Robinson:2005pd} 
	S.~P.~Robinson and F.~Wilczek,
	``A Relationship between Hawking radiation and gravitational anomalies,''
	Phys.\ Rev.\ Lett.\  {\bf 95}, 011303 (2005)
	[gr-qc/0502074].
	
	\bibitem{Iso:2006wa} 
	S.~Iso, H.~Umetsu and F.~Wilczek,
	``Hawking radiation from charged black holes via gauge and gravitational anomalies,''
	Phys.\ Rev.\ Lett.\  {\bf 96}, 151302 (2006)
	[hep-th/0602146].
	
	\bibitem{Majhi:2011yi} 
	B.~R.~Majhi,
	``Quantum Tunneling in Black Holes,''
	PhD Thesis, University of Calcutta, India,
	arXiv:1110.6008 [gr-qc].
	
	\bibitem{Adhikari:2017gyb} 
	A.~Adhikari, K.~Bhattacharya, C.~Chowdhury and B.~R.~Majhi,
	``Fluctuation-dissipation relation in accelerated frames,''
	Phys.\ Rev.\ D {\bf 97}, no. 4, 045003 (2018)
	[arXiv:1707.01333 [gr-qc]].
	
	\bibitem{Das:2019aii} 
	A.~Das, S.~Dalui, C.~Chowdhury and B.~R.~Majhi,
	``Conformal vacuum and the fluctuation-dissipation theorem in a de Sitter universe and black hole spacetimes,''
	Phys.\ Rev.\ D {\bf 100}, no. 8, 085002 (2019)
	[arXiv:1902.03735 [gr-qc]].
	
	\bibitem{Caldeira}
	A.~O.~Caldeira and A.~J.~Leggett,
	``Path integral approach to quantum Brownian motion,''
	Physica {\bf 121A}, 587 (1983).
	
	\bibitem{Unruh:1989}
	W.~G.~Unruh and W.~H.~Zurek,
	``Reduction of a Wave Packet in Quantum Brownian Motion,''
	Phys.\ Rev.\ D {\bf 40} (1989) 1071.
	
	\bibitem{Raine:1991}
	D.~J.~Raine, D.~W.~Sciama and P.~G.~Grove,
	``Does a uniformly accelerated quantum oscillator radiate?,''
	Proc.\ R.\ Soc.\ Land.\ {\bf A 435} (1991) 205-215.
	
	\bibitem{Hinterleitner:1993}
	F.~Hinterleitner,
	``A Model detector for Hawking radiation from a Schwarzschild black hole,''
	Fortsch.\ Phys.\  {\bf 43} (1995) 207.
	
	\bibitem{Kim:1997}
	H.~C.~Kim and J.~K.~Kim,
	``Radiation from a uniformly accelerating harmonic oscillator,''
	Phys.\ Rev.\ D {\bf 56} (1997) 3537
	[gr-qc/9704030].
	
	\bibitem{Kim:1998}
	H.~C.~Kim,
	``Quantum field and uniformly accelerated oscillator,''
	Phys.\ Rev.\ D {\bf 59} (1999) 064024
	[gr-qc/9808056].
	
	
	\bibitem{Deser:1976yx} 
	S.~Deser, M.~J.~Duff and C.~J.~Isham,
	``Nonlocal Conformal Anomalies,''
	Nucl.\ Phys.\ B {\bf 111}, 45 (1976).
	
	\bibitem{Duff:1977ay} 
	M.~J.~Duff,
	``Observations on Conformal Anomalies,''
	Nucl.\ Phys.\ B {\bf 125}, 334 (1977).
	
	\bibitem{Polyakov:1981rd} 
	A.~M.~Polyakov,
	``Quantum Geometry of Bosonic Strings,''
	Phys.\ Lett.\ B {\bf 103}, 207 (1981).
	
	\bibitem{Christensen:1977jc} 
	S.~M.~Christensen and S.~A.~Fulling,
	``Trace Anomalies and the Hawking Effect,''
	Phys.\ Rev.\ D {\bf 15}, 2088 (1977).
	
	\bibitem{Bardeen:1984pm} 
	W.~A.~Bardeen and B.~Zumino,
	``Consistent and Covariant Anomalies in Gauge and Gravitational Theories,''
	Nucl.\ Phys.\ B {\bf 244}, 421 (1984).
	
	\bibitem{AlvarezGaume:1984dr}
	L.~Alvarez-Gaume and P.~H.~Ginsparg,
	``The Structure of Gauge and Gravitational Anomalies,''
	Annals Phys.\  {\bf 161} (1985) 423
	Erratum: [Annals Phys.\  {\bf 171} (1986) 233].
	
	\bibitem{AlvarezGaume:1983ig} 
	L.~Alvarez-Gaume and E.~Witten,
	``Gravitational Anomalies,''
	Nucl.\ Phys.\ B {\bf 234}, 269 (1984).
	
	\bibitem{Leutwyler:1984nd} 
	H.~Leutwyler,
	``Gravitational Anomalies: A Soluble Two-dimensional Model,''
	Phys.\ Lett.\  {\bf 153B}, 65 (1985)
	Erratum: [Phys.\ Lett.\  {\bf 155B}, 469 (1985)].
	
	\bibitem{Frolov1} 
	 V.~P.~Frolov and A.~Zelnikov, ``Introduction to black hole physics,'' Oxford University Press, (2011).
	
\bibitem{Umetsu:2009ra} 
  K.~Umetsu,
  ``Hawking Radiation from Kerr-Newman Black Hole and Tunneling Mechanism,''
  Int.\ J.\ Mod.\ Phys.\ A {\bf 25}, 4123 (2010)
  [arXiv:0907.1420 [hep-th]].
  
  \bibitem{Banerjee:2010ye} 
  R.~Banerjee, B.~R.~Majhi, S.~K.~Modak and S.~Samanta,
  ``Killing Symmetries and Smarr Formula for Black Holes in Arbitrary Dimensions,''
  Phys.\ Rev.\ D {\bf 82}, 124002 (2010)
  [arXiv:1007.5204 [gr-qc]].
	
	\bibitem{book2}
	N.~D.~Birrell and P.~C.~W.~Davies, 
	``Quantum Fields in Curved Space,'' 
	Cambridge University Press, Cambridge, England, (1982).
	
	\bibitem{Banerjee:2007qs} 
	R.~Banerjee and S.~Kulkarni,
	``Hawking radiation and covariant anomalies,''
	Phys.\ Rev.\ D {\bf 77}, 024018 (2008)
	[arXiv:0707.2449 [hep-th]].
	
	\bibitem{Banerjee:2007uc} 
	R.~Banerjee and S.~Kulkarni,
	``Hawking radiation, effective actions and covariant boundary conditions,''
	Phys.\ Lett.\ B {\bf 659}, 827 (2008)
	[arXiv:0709.3916 [hep-th]].
	
	\bibitem{Banerjee:2008wq} 
	R.~Banerjee and S.~Kulkarni,
	``Hawking Radiation, Covariant Boundary Conditions and Vacuum States,''
	Phys.\ Rev.\ D {\bf 79}, 084035 (2009)
	[arXiv:0810.5683 [hep-th]].
	
	\bibitem{Banerjee:2008sn} 
	R.~Banerjee and B.~R.~Majhi,
	``Connecting anomaly and tunneling methods for Hawking effect through chirality,''
	Phys.\ Rev.\ D {\bf 79}, 064024 (2009)
	[arXiv:0812.0497 [hep-th]].

    \bibitem{Padmanabhan:1987rq} 
	T.~Padmanabhan and T.~P.~Singh,
	``Response of an Accelerated Detector Coupled to the Stress - Energy Tensor,''
	Class.\ Quant.\ Grav.\  {\bf 4}, 1397 (1987).
	
\bibitem{Frolov2}
V.~P.~Frolov and I.~Novikov, `` Black hole physics: Basic concepts and new developments,'' Kluwer academic publishers (1998).
	
	



	
	
	
	
	
	
	
	
	
	
	
	
		
	
	
	
	
	
	
	
	
		
	
		

	

	
	
\end{thebibliography}
\end{document}